\documentclass[showpacs,preprintnumbers]{revtex4-1}
\usepackage{amsmath,mathtools}
\usepackage[titletoc,title]{appendix}
\usepackage[dvips]{epsfig}

\def \be{\begin{equation}}
\def \ee{\end{equation}}
\def \bdm{\begin{eqnarray}}
\def \edm{\end{eqnarray}}
\begin{document}
\preprint{Submitted to Physics of Plasmas}
\title{Magnetic field line random walk in two-dimensional dynamical turbulence}
\author{J. F. Wang$^{1}$, G. Qin$^{2}$, Q. M. Ma$^{1}$, T. Song$^{1}$, S. B. Yuan$^{1}$}
\affiliation{$^{1}$Research Department of Biomedical Engineering, Institute of Electrical Engineering, Chinese Academy of Science, Beijing 100190, China; wangjunfang@mail.iee.ac.cn\\
$^{2}$School of Science, Harbin Institute of Technology, Shenzhen 518055, China; qingang@hit.edu.cn}
\date{\today}% It is always \today, today, but any date may be explicitly specified
\begin{abstract}
The field line random walk (FLRW) of magnetic turbulence is one of the
important topics in
plasma physics and astrophysics. In this article by using
the field line tracing method mean square displacements
(MSD) of FLRW is calculated in all possible
length scales for pure two-dimensional turbulence
with damping dynamical model.
We demonstrate that in order to describe FLRW
with damping dynamical model a new dimensionless quantity
$R$ is needed to be introduced. In different length scales
dimensionless MSD shows different relationship with the dimensionless quantity $R$.
Although temporal effect impacts MSD
of FLRW and even changes regimes of FLRW, it dose not affect
the relationship between the dimensionless MSD and dimensionless
quantity $R$ in all possible length scales.
\end{abstract}
\maketitle
\section{Introduction}
Field line random walk (FLRW) or field line wandering of magnetic  turbulence is one of the major problems
in the study of the galactic confinement of cosmic rays
(see, e.g., Jokipii \& Parker 1969), particle acceleration
at perpendicular shocks (see, e.g., Giacalone \& Jokipii 1996),
heat conduction by electrons (see, e.g., Chandran \& Cowley 1998; Lazarian 2006),
and many other astrophysical applications.
It has been demonstrated in several articles that FLRW directly
affects the transport of charged particles
(see, e.g., Skilling et al 1974;
Giacalone \& Jokipii 1999;
K\'ota \& Jokipii 2000; Matthaeus et al 2003; Shalchi \& Kourakis 2007a;
Shalchi \& Kourakis 2007b; Shalchi et al 2007).
So the knowledge of field line wandering is also important
for understanding the properties of energetic charged particles
propagation through the interplanetary or interstellar system
and ion diffusion in fusion devices (see, e.g., K\'ota \& Jokipii 2000;
Schlickeiser 2002; Matthaeus et al 2003;
Webb et al 2006; Qin 2007; Shalchi et al 2007; Shalchi \& Kourakis 2007c;
Tautz et al 2008; Shalchi et al 2009;
Weinhorst \& Shalchi 2010; Webb et al 2009;
Buffie \& Shalchi 2012; Qin \& Zhang 2014).

According to observations the total magnetic field of
the magnetized plasma is usually considered as
the superposition of a mean magnetic field $B_0$ and
a turbulent component $\delta \vec{B}$. Such magnetic field
configuration can be found in the solar system,
the Galactic space and fusion devices, e.g., Tokamaks.
Since the middle of the last century various analytical theories
have been developed to describe field line wandering
in such magnetic field configuration.
The first one is the quasilinear
theory which can be understood
as a first-order perturbation theory (see, Jokipii 1966).
However, the quasilinear
theory only works in pure slab (one-dimensional)
magnetic turbulence for parallel transport
with flat spectrum and
for perpendicular diffusion if pitch-angle scattering
is suppressed and if the Kubo number is small (Shalchi 2015).
For two-dimensional, quasi-3D (three-dimensional) and 3D
magnetic turbulence the quasilinear
theory is problematic to compute
FLRW and nonlinear description is essential (Shalchi 2009).
So Matthaeous et al (1995) developed a nonlinear analytical
theory for FLRW based on
the diffusion assumption.
But the superdiffusive regime of field line wandering was soon found
by computer simulations
(see, e.g., Zimbardo et al 1995; Pommois et al 1999).
Therefore, the diffusion theory of Matthaeous et al (1995) had to be
extended to describe nondiffusion regime
(see, Shalchi \& Kourakis 2007a).
Whereafter, some articles demonstrated
that the energy range index of the turbulence
spectrum determines whether random walk on magnetic field lines is diffusive or not
(see, e.g., Shalchi \& Kourakis 2007b, Shalchi \& Weinhost 2009),
which
was confirmed by Shalchi \& Qin (2010).
Furthermore, it was found by analytical investigations
that the spectral anisotropy
is another important factor affecting field line wandering
(see Weinhorst \& Shalchi 2010).
But it is not clear
that the most realistic regimes
of field line wandering is superdiffusive or diffusive.
With regard to the confusion about the regimes of field line wandering,
it was clarified by the in-depth study
of Shalchi (2011) that random walk on magnetic field lines complies
with different transport regimes
for different length scales.

The papers listed above only explored FLRW
of magnetostatic turbulence. But the real magnetic turbulence
should be time dependent, and it has been found that
dynamical turbulence effects have an important influence
on FLRW and transport
of energetic charged particle (see Shalchi 2009).
For slab turbulence with plasma wave model Shalchi et al (2007)
obtained a classic diffusive result for random walk on magnetic field lines.
Later, with regard to damping model
and random sweeping model Shalchi (2010a) found that
dynamical models are less restrictive concerning the allowed
energy range index than magnetostatic case for transport regime
of field line wandering.
And for pure two-dimensional turbulence
with damping dynamical model
and nonlinear anisotropic
dynamical turbulence model Guest \& Shalchi (2012)
found that both energy range index $q$
and the choice of the
dynamical models have an strong impact
on the field line wandering.
The articles mentioned above only explored the features of FLRW
in large spatial scale.
Shalchi (2011) investigated the properties
of FLRW for pure two-dimensional magnetostatic turbulence
in all possible length scales, i.e., inertial range, energy range
and even the range larger than the box scale.
But the influence of temporal effect
on field line wandering in all possible spatial length scales is
an open topic, and the implicit dimensionless quantity
controlling FLRW of dynamical turbulence is also
an unresolved issue.
The purpose of this article is to explore the these problems.
For mathematical tractability in this paper we only investigate
the impact of the simple damping model
with constant temporal factor
on FLRW of pure two-dimensional turbulence.

The organization of the article is as follows.
In Section II we show the
pure two-dimensional magnetic turbulence mode.
In Section III we describe the theory of FLRW
developed by Shalchi \& Kourakis (2007a).
In Section IV we extend the basic formulas of FLRW
in all possible spatial length scales
for pure two-dimensional magnetostatic turbulence (Shalchi 2011)
to dynamical turbulence.
In Section V, Section VI and Section VII we derive
the specific analytical formulas of mean square
displacement of field line wandering in different
length scales.
Section VIII shows summary and conclusion.

\section{Pure two-dimensional turbulence model}
In this article we explore field line wandering
of pure two-dimensional dynamical turbulence.
And the turbulent magnetic field component
$\delta \vec{B}(\vec{x})$ is oriented perpendicular with respect
to the mean magnetic field, i.e., $\delta B_z=0$.
The tensor of the 2D (two-dimensional) magnetic
turbulence has the following form (see Shalchi 2009)
\begin{equation}
P_{lm}^{2D}(\vec{k})=g^{2D}(k_\bot)\frac{\delta(k_\parallel)}
{k_\bot}\left[\delta_{lm}-\frac{k_l k_m}{k^2}\right],
\hspace{1cm} l,m=x,y  \label{2D tensor}
\end{equation}
where $\delta_{lm}$ is the Kronecker delta
and $\delta (k_\parallel)$ is the Dirac delta.
And $g^{2D}(k_\bot)$ is the spectrum of the two-dimensional
modes adopted in Shalchi (2011)
\begin{equation}
g^{2D}(k_\bot)=\frac{D(s,q)}{2\pi}l_{2D}\delta B_{2D}^2
\begin{cases}
0,                     &   \text{$k_\bot < L_{2D}^{-1}$}\\
(k_\bot l_{2D})^q,     &   \text{$L_{2D}^{-1}<k_\bot<l_{2D}^{-1}$}\\
(k_\bot l_{2D})^{-s},  &   \text{$l_{2D}^{-1}<k_\bot$}.
\label{2D mode}
\end{cases}
\end{equation}
Here $l_{2D}$ denotes the turnover
or bendover scale, $L_{2D}$ stands for
box scale which denotes the largest scale of
the stochastic magnetic field system, $\delta B_{2D}$ is
the field strength of the two-dimensional modes,
$s$ is the inertial range spectral index,
and $q$ is the energy range spectral index.
The region between box scale and turnover scale
is usually called energy range, and it is
inertial range for $k_\bot > l_{2D}^{-1}$.
The normalization function is given by
\begin{equation}
D(s,q)=\frac{(q+1)(s-1)}{s+q}.
\end{equation}
Here Eq. (\ref{2D mode})
is correctly normalized for $q>-1$ and $s>1$.
In addition, we assume $l_{2D}\ll L_{2D}$
throughout this article.

\section{THEORY OF MAGNETIC FIELD LINE RANDOM WALK}
To study field line wandering of magnetic turbulence we have
to use the method of statistical physics.
And the mean square
displacement (MSD)
$\langle(\Delta x(z))^2\rangle$ of magnetic field line
is the most frequently
used physical quantity to describe the properties of turbulent
magnetic field,
where $\Delta x(z)=x(z)-x(0)$ is the cross-field distance
and $\langle \cdots \rangle$ is ensemble average operator.
In the most previous investigations, a power law
$\langle(\Delta x)^2 \rangle=\alpha \left|z\right|^\beta$
with positive constants $\alpha$ and $\beta$
is usually used
to distinguish between different transport regimes according to
different $\beta$, namely, $0<\beta<1$ for subdiffusion,
$\beta=1$ for diffusion, $1<\beta<2$ for superdiffusion,
and $\beta=2$ for ballistic process,
where $\left|z\right|$ is the absolute value of distance
along the mean magnetic field (see, e.g., Shalchi 2009).
In what follows, we introduce the well-known theory of FLRW
developed by Shalchi \& Kourakis (2007a).

The equation of the stochastic magnetic field line reads
\begin{equation}
\frac{dx}{dz}=\frac{\delta B_x (\vec{x},t)}{B_0},
\label{motion equation of field line}
\end{equation}
where $z$ and $x$ are the displacement parallel
and perpendicular to the
background magnetic field $\vec{B}_0$ respectively,
and $\delta B_x (\vec{x},t)$
denotes $x$-component
of $\delta \vec{B} (\vec{x},t)$.

From Eq. (\ref{motion equation of field line})
the mean square displacement of magnetic field line
can be written as
\begin{equation}
\langle (\Delta x)^2 \rangle =\frac{1}{B_0 ^2}
\int_0 ^z dz' \int_0 ^z dz'' \langle \delta B_x (\vec{x}(z'),t')
\delta B_x (\vec{x}(z''),t'') \rangle.
\end{equation}
By operating Fourier transformation on the correlation
tensor of magnetic turbulence
the latter equation can be rewritten as
\begin{equation}
\langle (\Delta x)^2 \rangle =\frac{2}{B_0 ^2}
\int d^3 k P_{xx}(\vec{k}, t)\int_0^z dz' (z-z')
\langle e^{i \vec{k}\cdot \Delta \vec{x}(z)} \rangle,
\end{equation}
here spatial and temporal homogeneous assumptions and
Corrsin independence hypothesis (see, Corrsin 1959) are used.
Taking the second derivative
of the latter equation over $z$ gives
\begin{equation}
\frac{d^2 \langle(\Delta x)^2\rangle }{dz^2}
=\frac{2}{B_0 ^2} \int d^3 kP_{xx}(\vec{k},t)
\langle e^{i\vec{k}\cdot\Delta\vec{x}(z)}\rangle.
\end{equation}
For mathematical tractability we assume all tensor
components of turbulence obey same temporal behavior.
Then we can obtain
$P_{xx}(\vec{k},t)=P_{xx}(\vec{k},0)\Gamma(\vec{k},t)$
with the static tensor components $P_{xx}(\vec{k},0)$
and dynamical correction function $\Gamma(\vec{k},t)$.
Thus, we can get
\begin{equation}
\frac{d^2 \langle (\Delta x)^2\rangle}{dz^2}
=\frac{2}{B_0 ^2} \int d^3 k P_{xx}(\vec{k},0)
\Gamma(\vec{k},t)\langle e^{i\vec{k}\cdot
\Delta\vec{x}(z)}\rangle.
\label{Original second order derivative equation with ensemble average}
\end{equation}

\section{FIELD LINE RANDOM WALK FOR PURE TWO-DIMENSIONAL
TURBULENCE WITH DAMPING MODEL}
In this section we extend the basic formulas of FLRW
in all possible spatial scales
for pure two-dimensional magnetostatic turbulence
(Shalchi 2011) to dynamical turbulence.
For the sake of simplicity we only use the simple
exponential decaying model for turbulent dynamical
effect, i.e., the damping model
\begin{equation}
\Gamma(\vec{k},t)=e^{-\gamma t}, \label{damping model}
\end{equation}
where the parameter $\gamma$
is the characteristic temporal factor (see Bieber et al. 1994).
For slab and two-dimensional turbulence
temporal factor $\gamma$ is usually set as
$\alpha v_A k$ with constant parameter
$\alpha$, Alfv\'en wave speed $v_A$ and wave number $k$
(see Bieber et al. 1994; Shalchi 2010a; Guest \& Shalchi 2012).
In this article
for mathematical tractability we assume
that the temporal factor $\gamma$
is a constant. We leave the case $\gamma=\alpha v_A k$
and other more complicated cases for future work.

By assuming the Gaussian distribution of the magnetic field line to evaluate the
characteristic function $\langle e^{i\vec{k}\cdot \Delta\vec{x}(z)}\rangle $,
Eq. (\ref{Original second order derivative equation with ensemble average})
can be rewritten as
\begin{equation}
\frac{d^2 \langle (\Delta x)^2\rangle }{dz^2}
=\frac{2}{B_0 ^2} \int d^3 k P_{xx}(\vec{k})
e^{-\gamma t}e^{-\frac{1}{2} \langle(\Delta x)^2
\rangle k_\bot ^2} e^{ik_\parallel z}.
\label{Original second order derivative equation with Gaussian distribution}
\end{equation}
where we assume that magnetic turbulence is axisymmetric
with respect to background magnetic field.

In this article we compute the mean square
displacement of stochastic magnetic field line
by employing the field line tracing method
(see Shalchi 2010a), which assumes that
an energetic charged particle or an aerocraft
moves along magnetic field line with constant
$z$-component speed $v=z/t$. By setting
$\sigma=\langle(\Delta x)^2\rangle$ and
substituting time $t$ with $z/v$ in
Eq.
(\ref{Original second order derivative equation with Gaussian distribution}),
one can find
\begin{equation}
\frac{d^2 \sigma}{dz^2}=\frac{2}{B_0 ^2}
\int d^3 k P_{xx}(\vec{k})e^{-\frac{\gamma}{v}z}
e^{-\frac{1}{2} \sigma k_\bot ^2} e^{ik_\parallel z}.
\label{second order derivative equation with tracing method}
\end{equation}

By using the tensor of the two-dimensional magnetic turbulence
(see Eq. (\ref{2D tensor})), from
Eq. (\ref{second order derivative equation with tracing method})
we can obtain the following equation
\begin{equation}
\frac{d^2\sigma}{dz^2}=\frac{2\pi}{B_0 ^2}
\int_{0}^{\infty} dk_\bot g^{2D}(k_\bot)
e^{-\frac{\gamma}{v}z} e^{-\frac{1}{2}
\sigma k_\bot ^2}.
\label{Second-order derivative equation of MSD}
\end{equation}

After multiplying Eq. (\ref{Second-order derivative equation of MSD})
by $d\sigma/dz$, integrating the result by parts yields
\begin{equation}
\begin{aligned}
\left(\frac{d\sigma}{dz}\right)^2
= &\frac{8\pi}{B_0 ^2} \int_{0}^{\infty} dk_\bot
g^{2D}(k_\bot)k_\bot ^{-2}-\frac{8\pi}{B_0 ^2}
e^{-\frac{\gamma}{v}z} \int_{0}^{\infty} dk_\bot
g^{2D}(k_\bot)k_\bot ^{-2} e^{-\frac{1}{2}
\sigma k_\bot ^2}  \\   %\nonumber
&-\frac{8\pi}{B_0 ^2}\frac{\gamma}{v}\int_{0}^{z} dz'
e^{-\frac{\gamma}{v}z'} \int_{0}^{\infty} dk_\bot
g^{2D}(k_\bot)k_\bot ^{-2} e^{-\frac{1}{2}
\sigma k_\bot ^2},
\label{(dMSD/dz)2 without simplification}
\end{aligned}
\end{equation}
where the original conditions
$\sigma(z=0)=0$ and $(d\sigma / dz)(z=0)=0$ are used.
The first term on the right hand side of
Eq. (\ref{(dMSD/dz)2 without simplification})
is time-independent, but the second and third terms
are affected by temporal effect.
If setting $\gamma=0$
Eq. (\ref{(dMSD/dz)2 without simplification})
is simplified as the magnetostatic turbulence result
(see Eq. (16) in Shalchi 2011).

Substituting the formula
of two-dimensional modes (see, Eq. (\ref{2D mode}))
into Eq. (\ref{(dMSD/dz)2 without simplification}),
and then using the transformation $y=l_{2D}k_\bot$,
finally one can obtain
\begin{eqnarray}
%\begin{aligned}
\left(\frac{d\sigma}{dz}\right)^2&=&
4D(s,q)\frac{\delta B_{2D}^2}{B_0 ^2}l_{2D}^2
\left(\frac{1-\xi^{q-1}}{q-1}+\frac{1}{s+1}\right) \nonumber\\
&{}&-2D(s,q)\frac{\delta B_{2D}^2}{B_0 ^2}l_{2D}^2
e^{-\frac{\gamma}{v}z}\nonumber\\
&{}&\times\left\{\rho^{1-q}
\left[\Gamma\left(\frac{q-1}{2},\rho^2 \xi^2\right)
-\Gamma\left(\frac{q-1}{2},\rho^2 \right)\right]
+\rho^{s+1}\Gamma\left(-\frac{s+1}{2},\rho^2 \right)\right\}
\label{(dMSD/dz)^2}\\
&{}&-2D(s,q)\frac{\delta B_{2D}^2}{B_0 ^2}l_{2D}^2
\frac{\gamma}{v} \nonumber\\
&{}&\times\int_{0}^{z}dz' e^{-\frac{\gamma}{v}z'}
\left\{\rho^{1-q}\left[\Gamma\left(\frac{q-1}{2},
\rho^2 \xi^2\right)
-\Gamma\left(\frac{q-1}{2},\rho^2 \right)\right]
+\rho^{s+1}\Gamma\left(-\frac{s+1}{2},\rho^2
\right)\right\}, \nonumber
%\end{aligned}
\end{eqnarray}
where the following formulas are used
(see Gradshteyn \& Ryzhik 2007)
\begin{eqnarray}
\int_{\xi}^{1} y^{q-2}dy&=&\frac{1-\xi ^{q-1}}{q-1}, \\
\int_{1}^{\infty} y^{-s-2}dy&=&\frac{1}{s+1},    \\
\int_{\xi}^{1} y^{q-2}e^{-\rho ^2 y^2}dy
&=&\frac{1}{2}\rho ^{1-q}\left[\Gamma
\left(\frac{q-1}{2},\rho^2 \xi^2\right)
-\Gamma\left(\frac{q-1}{2},\rho^2 \right)\right], \\
\int_{1}^{\infty} y^{-s-2}e^{-\rho ^2 y^2}dy
&=&\frac{1}{2}\rho ^{s+1}\Gamma\left(-\frac{s+1}{2},
\rho^2 \right).
\end{eqnarray}
Here $\Gamma(\nu,z)$ is the upper incomplete gamma function,
the parameters $\xi=l_{2D}/L_{2D}\ll 1$, $\rho^2
=\sigma/(2l_{2D}^2)$
and $\rho^2 \xi^2=\sigma/(2L_{2D}^2)$ are also used.

Upon two order differentiation on Eq. (\ref{(dMSD/dz)^2})
with respect to $z$,
we can get the ordinary differential equation of the mean
square displacement which describes
the properties of FLRW in all length scales
\begin{eqnarray}
%\begin{aligned}
\frac{d^2\sigma}{dz^2}&=&-\frac{D(s,q)}{2}
\frac{\delta B_{2D}^2}{B_0 ^2}e^{-\frac{\gamma}{v}z}
\Bigg\{\frac{1-q}{2}\left[\Gamma\left(\frac{q-1}{2},
\rho^2 \xi^2\right)-\Gamma\left(\frac{q-1}{2},
\rho^2 \right)\right]\left(\frac{2l_{2D}^2}{\rho^2}\right)
^{\frac{1+q}{2}}\nonumber\\
&{}&+\left(e^{-\rho^2}-e^{-\rho^2\xi^2}\xi^{q-1}\right)
\frac{2l_{2D}^2}{\sigma}+\frac{s+1}{2}
\left(\frac{2l_{2D}^2}{\sigma}\right)^{\frac{1-s}{2}}
\Gamma\left(-\frac{s+1}{2},\rho^2\right)
-e^{-\rho^2}\frac{2l_{2D}^2}{\sigma}\Bigg\}.
%\end{aligned}
\end{eqnarray}
In the following sections,
we consider the special cases in different length scales.

\section{ANALYTICAL FORMULAS OF FIELD LINE WANDERING IN THE RANGE
$\boldsymbol{\sigma \ll 2l_{2D}^2 \ll 2L_{2D}^2}$}
In the range $\sigma \ll 2l_{2D}^2 \ll 2L_{2D}^2$
we can see that the conditions $\rho^2\ll 1$
and $\rho^2\xi^2\ll 1$ need to be satisfied.
\subsection{The features of field line random walk in the range
$\boldsymbol{\sigma \ll 2l_{2D}^2 \ll 2L_{2D}^2}$}
By using the conditions $\rho^2\ll 1$ and
$\rho^2\xi^2\ll 1$ we can derive from Eq. (\ref{(dMSD/dz)^2})
\begin{eqnarray}
%\begin{aligned}
\left(\frac{d\sigma}{dz}\right)^2
	&\approx &4D(s,q)\frac{\delta B_{2D}^2}{B_0 ^2}l_{2D}^2
\left(\frac{1-\xi^{q-1}}{q-1}+\frac{1}{s+1}\right)
-4D(s,q)\frac{\delta B_{2D}^2}
	{B_0 ^2}l_{2D}^2e^{-\frac{\gamma}{v}z}\times\nonumber\\
	&{}&\left[\frac{1-\xi^{q-1}}{q-1}
+\rho^2\frac{\xi^{q+1}-1}{q+1}
+\frac{\rho^{s+1}}{2}\Gamma\left(-\frac{s+1}{2}
\right)+\frac{1}{s+1}+\frac{\rho^2}{1-s}\right]
\label{Original (dMSD/dz)^2 in small scale}\\
	&{}&-4D(s,q)\frac{\delta B_{2D}^2}{B_0 ^2}l_{2D}^2
\frac{\gamma}{v}\int_{0}^{z}dz' e^{-\frac{\gamma}{v}z'}
\Bigg[\frac{1-\xi^{q-1}}{q-1}
+\rho^2\frac{\xi^{q+1}-1}{q+1}
+\frac{\rho^{s+1}}{2}\Gamma\left(-\frac{s+1}{2}\right)\nonumber\\
	&{}&+\frac{1}{s+1}+\frac{\rho^2}{1-s}\Bigg],\nonumber
%\end{aligned}
\end{eqnarray}
where the following formulas are used
(see Abramowitz \& Stegun 1974)
\begin{equation} \label{Limit form of Gamma function}
\begin{split}
&\Gamma(\nu,z\gg 1)\approx z^{\nu-1}e^{-z}\rightarrow 0, \\
&\Gamma(\nu,z\ll 1)\approx \Gamma(\nu)-\frac{z^\nu}{\nu}+\frac{z^{\nu +1}}{\nu+1}. \\
\end{split}
\end{equation}

By integrating by parts,
Eq. (\ref{Original (dMSD/dz)^2 in small scale})
can be simplified as
\begin{equation}
\left(\frac{d\sigma}{dz}\right)^2\approx 2D(s,q)
\frac{\delta B_{2D}^2}{B_0 ^2} \left(\frac{1}{q+1}
+\frac{1}{s-1}\right)\int_{0}^{z}d\sigma
e^{-\frac{\gamma}{v}z'}.
\label{(dMSD/dz)^2 in small scale }
\end{equation}
To differentiate Eq. (\ref{(dMSD/dz)^2 in small scale })
over z, one can obtain
\begin{equation}
\frac{d^2\sigma}{dz^2}\approx\frac{\delta B_{2D}^2}{B_0 ^2}
e^{-\frac{\gamma}{v}z}.
\label{Second derivative equation in small scale}
\end{equation}
By setting the original conditions as
$(d\sigma/dz)(z=0)=0$ and $\sigma(z=0)=0$ as in above section,
from Eq. (\ref{Second derivative equation in small scale})
mean square displacement for the case
$\sigma \ll 2l_{2D}^2 \ll 2L_{2D}^2$ can be found
\begin{equation}
\sigma\approx\frac{\delta B_{2D}^2}{B_0 ^2}\frac{v}{\gamma}
\left[z+\frac{v}{\gamma}
\left(e^{-\frac{\gamma}{v}z}-1\right)\right].
\label{MSD in small scale}
\end{equation}

Note that FLRW in the range
$\sigma \ll 2l_{2D}^2 \ll 2L_{2D}^2$
is no longer the simple quadratic function
of the parallel position $z$ (ballistic process)
as in the magnetostatic case
(see Shalchi 2011). In addition,
we find that the energy range index $q$ and
the inertial range index $s$ have no any influence on
the features of field line wandering.
In the following, we explore the properties
of FLRW for some special cases in the range
$\sigma \ll 2l_{2D}^2 \ll 2L_{2D}^2$.

For the weak dynamical limit $\gamma\ll 1$
corresponding to the quasi-magnetostatic case, Eq.
(\ref{MSD in small scale}) is simplified as
\begin{equation}
\sigma\approx\frac{1}{2}\frac{\delta B_{2D}^2}{B_0 ^2}z^2.
\end{equation}
The latter equation is the same as
the magnetostatic result (see Eq. (26) in Shalchi 2011).

If temporal factor $\gamma$ is a nonzero value,
from Eq. (\ref{MSD in small scale})
we can get following equation
in the interval $0<z\ll v/\gamma$
\begin{equation}
\sigma\approx\frac{1}{2}\frac{\delta B_{2D}^2}{B_0 ^2}z^2.
\label{MSD of first range in small scale}
\end{equation}

The latter equation is also identical with
the magnetostatic case. So we can see that
the temporal effect is negligible in the subrange
$0<z\ll v/\gamma$ regardless
of the strength of temporal effect.
But if temporal effect is strong enough,
the condition $v/\gamma \ll l_{2D}$ can be satisfied,
from Eq. (\ref{MSD in small scale})
one can obtain the following formulas
in the subrange $v/\gamma \ll z\ll l_{2D}$
\begin{eqnarray}
\kappa_{FLT}&\approx &\frac{1}{2}
\frac{\delta B_{2D}^2}{B_0 ^2}\frac{v}{\gamma},
\label{Diffusion coefficient for second range in small scale}   \\
\sigma &\approx & \frac{\delta B_{2D}^2}{B_0 ^2}
\frac{v}{\gamma}\left(z-\frac{v}{\gamma}\right).
\label{MSD of second range in small scale}
\end{eqnarray}
Eqs. (\ref{Diffusion coefficient for second range in small scale})
and (\ref{MSD of second range in small scale})
show that the regime of FLRW is diffusive.
%in the subrange $v/\gamma \ll z\ll l_{2D}$.

In summary, if temporal effect is strong enough,
the range $0<z\ll l_{2D}$ can been split into
two subrange: $0<z\ll v/\gamma \ll l_{2D}$ and
$v/\gamma \ll z\ll l_{2D}$. In the first subrange
dynamical effect can be neglected and magnetic
field line wandering presents ballistic process
regardless of the strength of temporal effect.
But in the second subrange the regime of FLRW
can be transformed from ballistic into
diffusive by temporal effect.
The stronger temporal effect, i.e., the larger factor
$\gamma$, leads to the longer subrange
$v/\gamma \ll z\ll l_{2D}$
and the smaller diffusion coefficient. Therefore,
the temporal effect not only can reduce MSD but also
change the regimes of magnetic field line wandering.
However, for magnetostatic turbulence it is only
ballistic in the whole range $0<z\ll l_{2D}$.
If the temporal effect is very weak
so that $v/\gamma \gg l_{2D}$,
it is only ballistic in the whole range $0<z\ll l_{2D}$
as same as in the magnetostatic case.

\subsection{Dimensionless quantities in the range
$\boldsymbol{\sigma \ll 2l_{2D}^2 \ll 2L_{2D}^2}$}
Taking the nondimensionlizing operation on
Eq. (\ref{MSD in small scale}), we can obtain
\begin{equation}
\sigma'\approx\frac{1}{2}R^2\left(z'+e^{-z'}-1\right),
\label{Dimensionless equation in small scale}
\end{equation}
here the dimensionless quantities $z'=\gamma z/v$
and $\sigma'=\sigma/(2l_{2D}^2)$ are used.
And we can see that a new dimensionless quantity
occurs in the latter equation as follow
\begin{equation}
R=\frac{\delta B_{2D}}{B_0}\frac{v}{\gamma l_{2D}}.
\end{equation}
From Eq. (\ref{Dimensionless equation in small scale})
we can see that the dimensionless quantity
$R$ controls FLRW in the range
$\sigma \ll 2l_{2D}^2 \ll 2L_{2D}^2$.
In addition, the dimensionless mean square
displacement $\sigma'$ is proportional to the square
of the dimensionless quantities $R$.

By employing the same nondimensionalizing method as above,
the governing equation in the subrange
$0<z\ll  v/\gamma \ll l_{2D}$ (see
Eq. (\ref{MSD of first range in small scale}))
can be rewritten as
\begin{equation}
\sigma'\approx\frac{1}{4}R^2 z'^2,
\label{Dimensionless equation of first range in small scale}
\end{equation}
and the governing equation in the subrange
$ v/\gamma \ll z\ll l_{2D}$ (see
Eq. (\ref{MSD of second range in small scale}))
can be nondimensionlized as
\begin{equation}
\sigma'\approx\frac{1}{2}R^2\left(z'-1\right).
\label{Dimensionless equation of second range in small scale}
\end{equation}

From Eqs.
(\ref{Dimensionless equation of first range in small scale})
and
(\ref{Dimensionless equation of second range in small scale})
we can see that the dimensionless quantity $R$
controls the features of dimensionless mean square displacement
$\sigma'$ in the subranges $0<z\ll  v/\gamma \ll l_{2D}$
and $ v/\gamma \ll z\ll l_{2D}$, i.e.,
in the whole range $0<z\ll l_{2D}$.
Although it is ballistic
in the former subrange and diffusive in the latter subrange,
dimensionless mean square displacement $\sigma'$ is
always proportional to $R^2$ in the whole range
$0<z\ll l_{2D}$.
Therefore, for the case
$\sigma \ll 2l_{2D}^2 \ll 2L_{2D}^2$
%$0<z\ll l_{2D}$
the dynamical effect
of magnetic turbulence might have an impact on the regimes
of FLRW, but it does not make any influence
on the relationship between $\sigma'$
and the new dimensionless quantity
$R$. In other words, although the dynamical effect
might change the regimes of FLRW from ballistic into diffusive,
it has no any impact on
the relations between $\sigma$ and the turbulence level
$\delta B_{2D}/B_0$.

\section{ANALYTICAL FORMULAS OF FIELD LINE WANDERING IN THE RANGE
$\boldsymbol{2l_{2D}^2\ll 2L_{2D}^2\ll \sigma}$}
In the range $2l_{2D}^2\ll 2L_{2D}^2\ll \sigma$,
i.e., in the range outside box size, the following
conditions must be satisfied
\begin{equation}
\rho^2=\frac{\sigma}{2l_{2D}^2}\gg 1, \hspace{1cm}
\rho^2 \xi^2=\frac{\sigma}{2L_{2D}^2}\gg 1,
\label{Requirement for energy range}
\end{equation}
here parameter $\xi=l_{2D}/L_{2D}\ll 1$ is used.

\subsection{The features in the range $\boldsymbol{2l_{2D}^2\ll 2L_{2D}^2\ll \sigma}$}
By employing Eqs. (\ref{Limit form of Gamma function})
and (\ref{Requirement for energy range}),
Eq. (\ref{(dMSD/dz)^2})
can be rewritten as
\begin{equation}
\begin{aligned}
\left(\frac{d\sigma}{dz}\right)^2=&4D(s,q)
\frac{\delta B_{2D}^2}{B_0 ^2}l_{2D}^2
\left(\frac{1-\xi^{q-1}}{q-1}+\frac{1}{s+1}\right)\\
&-2D(s,q)\frac{\delta B_{2D}^2}{B_0 ^2}
l_{2D}^2e^{-\frac{\gamma}{v}z}
\left[\rho^{-2}\xi^{q-3} e^{-\rho^2 \xi^2}
-\rho^{-2}e^{-\rho^2}+\rho^{-2}e^{-\rho^2}\right]\\
&-2D(s,q)\frac{\delta B_{2D}^2}{B_0 ^2}l_{2D}^2
\frac{\gamma}{v}\int_{0}^{z}dz' e^{-\frac{\gamma}{v}z'}
\left[\rho^{-2}
\xi^{q-3}e^{-\rho^2 \xi^2}-\rho^{-2}e^{-\rho^2}
+\rho^{-2}e^{-\rho^2}\right].
%\label{41}
\end{aligned}
\end{equation}
From Eq. (\ref{Requirement for energy range})
we can find that $e^{-\rho^2 \xi^2}$ and
$e^{-\rho^2}$ all tend to zero in the range
$2l_{2D}^2\ll 2L_{2D}^2\ll \sigma$.
Then the latter equation can be simplified as
\begin{equation}
\left(\frac{d\sigma}{dz}\right)^2\approx
4D(s,q)\frac{\delta B_{2D}^2}{B_0 ^2}l_{2D}^2
\left(\frac{1-\xi^{q-1}}{q-1}+\frac{1}{s+1}\right).
\label{(dMSD/dz)^2 for q>1 in energy range}
\end{equation}
Since $\xi^{q-1}\ll 1$ for $q>1$, Eq.
(\ref{(dMSD/dz)^2 for q>1 in energy range})
becomes
\begin{equation}
\left(\frac{d\sigma}{dz}\right)^2\approx 4D(s,q)
\frac{\delta B_{2D}^2}{B_0 ^2}l_{2D}^2
\left(\frac{1}{q-1}+\frac{1}{s+1}\right).
\end{equation}
By using the definition of diffusion coefficient
of field line $\kappa_{FLT}=(1/2) \ (d\sigma/dz)$,
we obtain
\begin{equation}
\kappa_{FLT}\approx\sqrt{\frac{(q+1)(s-1)}{(q-1)(s+1)}}l_{2D}
\frac{\delta B_{2D}}{B_0}.
\label{Diffusion coefficient of q>1 in energy range}
\end{equation}
Since $\xi^{q-1}\gg 1$ for $-1<q<1$, by using the
same method as above, from Eq.
(\ref{(dMSD/dz)^2 for q>1 in energy range})
we can deduce the diffusion coefficient as follow
\begin{equation}
\kappa_{FLT}\approx\sqrt{\frac{(q+1)(s-1)}{(\boldsymbol{s+q})(1-q)}}
l_{2D}\frac{\delta B_{2D}}{B_0}
\left(\frac{L_{2D}}{l_{2D}}\right)^{\frac{1-q}{2}}.
\label{Diffusion coefficient of -1<q<1 in energy range }
\end{equation}

We can find that Eqs.
(\ref{Diffusion coefficient of q>1 in energy range})
and (\ref{Diffusion coefficient of -1<q<1 in energy range })
are perfectly identical
with Eqs. (35) and (36) in the article of Shalchi (2011)
which are from magnetostatic model, respectively.
Therefore, we can see that
in the range larger than the box size the diffusion
coefficients are independent of the temporal effect
irrespective of the strength of temporal effect.
For this result we can obtain
some qualitative explanation from
Eq. (\ref{(dMSD/dz)2 without simplification}).
The first term on the right hand side of
Eq. (\ref{(dMSD/dz)2 without simplification})
is independent of temporal effect,
while the second term and third terms
are related to time $t$, that is, related to position $z$.
For the limit $\sigma(z)\rightarrow \infty$ corresponding to
$z\rightarrow \infty$, then $e^{-vz/\gamma}\rightarrow 0$,
so the second and third terms of Eq.
(\ref{(dMSD/dz)2 without simplification})
can be ignored in comparison with the first term.
That is, only the first term in Eq.
(\ref{(dMSD/dz)2 without simplification})
is remained for $z\rightarrow \infty$.
Therefore, for $z\rightarrow \infty$ the regime of
particle transport tends to the magnetostatic result,
i.e., diffusion (see Shalchi 2011).

\subsection{Dimensionless parameters in the range $\boldsymbol{2l_{2D}^2\ll 2L_{2D}^2\ll \sigma}$}
By using the same method as in the subsection V.B
and employing the dimensionless quantity
$R=(\delta B_{2D} v)/(B_0 \gamma l_{2D})$, from Eqs.
(\ref{Diffusion coefficient of q>1 in energy range})
and (\ref{Diffusion coefficient of -1<q<1 in energy range })
we can get the following dimensionless equations
\begin{eqnarray}
\kappa'_{FLT}&\approx &\frac{1}{2}\sqrt{\frac{(q+1)(s-1)}{(q-1)(s+1)}}R,
\hspace{2.3cm} \text{for} \hspace{0.3cm} q>1
\label{dimensionless diffusion coefficient for q>1 in large scale}    \\
\kappa'_{FLT}&\approx &\frac{1}{2}\sqrt{\frac{(q+1)(s-1)}{(1-q)(s+q)}}R
\xi^{(q-1)/2}, \hspace{1cm} \text{for} \hspace{0.3cm} -1<q<1
\label{dimensionless diffusion coefficient for 1-<q<1 in large scale}
\end{eqnarray}
where the dimensionless diffusion coefficient is defined as
$\kappa'_{FLT}=(1/2)(d\sigma'/dz')$.
From Eqs.
(\ref{dimensionless diffusion coefficient for q>1 in large scale})
and
(\ref{dimensionless diffusion coefficient for 1-<q<1 in large scale})
we can see that the dimensionless diffusion coefficient
$\kappa'_{FLT}$ is proportional to dimensionless parameter $R$
for any allowed value of $q$
in the range $2l_{2D}^2\ll 2L_{2D}^2\ll \sigma$.
In other words, the diffusion coefficient $\kappa_{FLT}$
is proportional to $\delta B_{2D}/B_0$
in the range $2l_{2D}^2\ll 2L_{2D}^2\ll \sigma$
regardless of the energy index $q$.

\section{ANALYTICAL FORMULAS OF FIELD LINE WANDERING IN THE RANGE
$\boldsymbol{2l_{2D}^2\ll \sigma\ll 2L_{2D}^2}$}
We can obtain the conditions
$\rho^2 =\sigma/(2l_{2D}^2)\gg 1$ and $\rho^2 \xi^2
=\sigma/(2L_{2D}^2)\ll 1$ in the range
$2l_{2D}^2\ll \sigma\ll 2L_{2D}^2$.
By using Eq.
(\ref{Limit form of Gamma function}) we can simplify Eq.
(\ref{(dMSD/dz)^2}) as follow
\begin{equation}
\begin{aligned}
\left(\frac{d\sigma}{dz}\right)^2
\approx &4D(s,q)\frac{\delta B_{2D}^2}{B_0 ^2}
l_{2D}^2 \left(\frac{1-\xi^{q-1}}{q-1}+\frac{1}{s+1}\right)\\
&-2D(s,q)\frac{\delta B_{2D}^2}{B_0 ^2}l_{2D}^2
e^{-\frac{\gamma}{v}z}\left[\rho^{1-q}\Gamma
\left(\frac{q-1}{2}\right)-\frac{2}{q-1}\xi^{q-1}
+\frac{2}{q+1}\rho^2 \xi^{q+1}+\rho^{-2}e^{-\rho^2}\right]\\
&-2D(s,q)\frac{\delta B_{2D}^2}{B_0 ^2}l_{2D}^2
\frac{\gamma}{v}
\int_{0}^{z}dz' e^{-\frac{\gamma}{v}z'}
\left[\rho^{1-q}\Gamma\left(\frac{q-1}{2}\right)
-\frac{2}{q-1}\xi^{q-1}+\frac{2}{q+1}\rho^2
\xi^{q+1}+\rho^{-2}e^{-\rho^2}\right].
\label{(dMSD/dz)^2 in intermediate range}
\end{aligned}
\end{equation}

\subsection{The special case $\boldsymbol{q>1}$}
For $q>1$ the second and third terms on the
right hand side of Eq.
(\ref{(dMSD/dz)^2 in intermediate range})
can be neglected comparing to the first term. Then
Eq. (\ref{(dMSD/dz)^2 in intermediate range})
can be simplified as
\begin{equation}
\left(\frac{d\sigma}{dz}\right)^2
\approx 4D(s,q)\frac{\delta B_{2D}^2}{B_0 ^2}
l_{2D}^2 \left(\frac{1}{q-1}+\frac{1}{s+1}\right).
\end{equation}
From the latter equation we can get the diffusion coefficient
of magnetic field line wandering as
\begin{equation}
\kappa_{FLT}\approx\sqrt{\frac{(q+1)(s-1)}{(q-1)(s+1)}}
\frac{\delta B_{2D}}{B_0}l_{2D},
\label{Diffusion coefficient with q>1 in intermediate range}
\end{equation}
which coincides with Eq. (29) in Shalchi (2011)
for the same special case.
Through nondimensionalizing of
Eq. (\ref{Diffusion coefficient with q>1 in intermediate range})
we can obtain
\begin{equation}
\kappa'_{FLT}\approx\frac{1}{2}\sqrt{\frac{(q+1)(s-1)}{(q-1)(s+1)}}R,
\end{equation}
where the dimensionless parameter
$R=(\delta B_{2D} v)/(B_0 \gamma l_{2D})$
occurs again.
Therefore, for this special case
the dimensionless diffusion coefficient
of FLRW is proportional to the new dimensionless parameter
$R$, or diffusion coefficient $\kappa_{FLT}$ is proportional
to turbulence level $\delta B_{2D}/B_0$.

\subsection{The special case $\boldsymbol{-1<q<1}$}
For $-1<q<1$ Eq. (\ref{(dMSD/dz)^2}) can be simplifies as
\begin{equation}
\begin{aligned}
\left(\frac{d\sigma}{dz}\right)^2
\approx &4D(s,q)\frac{\delta B_{2D}^2}{B_0 ^2}l_{2D}^2
\left(\frac{\xi^{q-1}}{1-q}+\frac{1}{s+1}\right)\\
&-2D(s,q)\frac{\delta B_{2D}^2}{B_0 ^2}l_{2D}^2
e^{-\frac{\gamma}{v}z}\left[\rho^{1-q}\Gamma
\left(\frac{q-1}{2}\right)-\frac{2}{q-1}\xi^{q-1}
+\frac{2}{q+1}\rho^2\xi^{q+1}\right]\\
&-2D(s,q)\frac{\delta B_{2D}^2}{B_0 ^2}l_{2D}^2
\frac{\gamma}{v}\int_{0}^{z}dz' e^{-\frac{\gamma}{v}z'}
\left[\rho^{1-q}\Gamma\left(\frac{q-1}{2}\right)
-\frac{2}{q-1}\xi^{q-1}+\frac{2}{q+1}\rho^2\xi^{q+1}\right].
\end{aligned}
\end{equation}

By taking the derivative of the latter equation over $z$
we can obtain
\begin{equation}
\frac{d^2\sigma}{dz^2}\approx \frac{D(s,q)}{2}\Gamma(\frac{q+1}{2})
\frac{\delta B_{2D}^2}{B_0 ^2}e^{-\frac{\gamma}{v}z}
\left(\frac{\sigma}{2l_{2D}^2}\right)^{-\frac{1+q}{2}}.
\label{Nonlinear Second order derivative equation of MSD}
\end{equation}

By nondimensionalizing the latter equation
we can obtain
\begin{equation}
\frac{d^2\sigma'}{dz'^2}\approx\frac{D(s,q)}{4}
\Gamma(\frac{q+1}{2})e^{-z'} \sigma'^{-\frac{1+q}{2}}R^2.
\label{Dimensionless second-order derivative equation}
\end{equation}

From the latter equation
we can see that the dimensionless quantity
$R$ controls the properties
of magnetic field line wandering
for the special case $-1<q<1$.
We set the following relationship
\begin{equation}
\sigma'= h(R)p(z').       \label{sigma'}
\end{equation}
Inserting Eq. (\ref{sigma'}) into Eq.
(\ref{Dimensionless second-order derivative equation})
we can obtain
\begin{equation}
h(R)\frac{d^2 p(z')}{dz'^2}\approx\frac{D(s,q)}{4}
\Gamma(\frac{q+1}{2}) e^{-z'} h(R)^{-\frac{1+q}{2}}
p(z')^{-\frac{1+q}{2}}R^2.
\end{equation}
Thus we can find the following formula
\begin{equation}
h(R)\propto R^2 h(R)^{-(1+q)/2},
\end{equation}
or
\begin{equation}
h(R)\propto R^{4/(3+q)}.
\label{h2}
\end{equation}
Therefore, from Eq. (\ref{sigma'}) we can obtain
\begin{equation}
\sigma'\approx h(R)p(z') \propto  R^{4/(3+q)}p(z').
\label{sigma and R}
\end{equation}
From the above discussion we can see that the relationship
(\ref{sigma and R})
is irrelevant to temporal effect.
So the dynamical effect has no impact
on the relationship between
the dimensionless mean square displacement
$\sigma'$ and the
dimensionless quantity $R$.

\subsection{Simplification of the governing equation for $\boldsymbol{-1<q<1}$ in the range
$\boldsymbol{2l_{2D}^2\ll \sigma\ll 2L_{2D}^2}$ }
Let's set $\sigma=g(z)e^{f(z)}$,
the governing equation  for $-1<q<1$ in the range
$2l_{2D}^2\ll \sigma\ll 2L_{2D}^2$
(see Eq.
(\ref{Nonlinear Second order derivative equation of MSD}))
can be simplified as
\begin{equation}
\begin{aligned}
\left[\frac{d^2g}{dz^2}+2\frac{dg}{dz}\frac{df}{dz}
+g\left(\frac{df}{dz}\right)^2+g\frac{d^2f}{dz^2}\right]
e^f=\frac{D(s,q)}{2^{(1-q)/2}}\frac{\delta B_{2D}^2}{B_0^2}
\Gamma(\frac{q+1}{2})l_{2D}^{1+q}
g^{-\frac{1+q}{2}}e^{-\frac{\gamma}{v}z}
e^{-\frac{1+q}{2}f}.
\end{aligned}
\end{equation}
By comparing the left-hand side of the latter equation
with the right-hand side,
we can get the following equations
\begin{equation}
f=-\frac{2}{3+q}\frac{\gamma}{v}z,
\label{F with z}
\end{equation}
and
\begin{equation}
\frac{d^2g}{dz^2}+2\frac{dg}{dz}\frac{df}{dz}
+g\left(\frac{df}{dz}\right)^2+g\frac{d^2f}{dz^2}
=\frac{D(s,q)}{2^{(1-q)/2}}\frac{\delta B_{2D}^2}{B_0^2}
\Gamma(\frac{q+1}{2})l_{2D}^{1+q}g^{-\frac{1+q}{2}}.
\label{Equation of f and g}
\end{equation}
From Eqs. (\ref{F with z}) and (\ref{Equation of f and g})
 we can obtain the
equation of $g(z)$ as
\begin{equation}
\frac{d^2g}{dz^2}-\frac{4}{3+q}\frac{\gamma}{v}
\frac{dg}{dz}+\left(\frac{2}{3+q}
\frac{\gamma}{v}\right)^2 g
=\frac{D(s,q)}{2^{(1-q)/2}}\frac{\delta B_{2D}^2}{B_0^2}
\Gamma(\frac{q+1}{2})l_{2D}^{1+q}g^{-\frac{1+q}{2}}.
\label{original Simplified first-order equation in main text}
\end{equation}
The latter equation can be rewritten as
\begin{equation}
\frac{d^2g}{dz^2}-2B\frac{dg}{dz}+B^2 g-Ag^{-\frac{1+q}{2}}=0
\label{Simplified first-order equation in main text}
\end{equation}
with
\begin{eqnarray}
A&=&\frac{D(s,q)}{2^{(1-q)/2}}\frac{\delta B_{2D}^2}{B_0^2}
\Gamma(\frac{q+1}{2})l_{2D}^{1+q},   \label{A}  \\
B&=&\frac{2}{3+q}\frac{\gamma}{v}.
\label{B}
\end{eqnarray}
By employing Eq. (\ref{F with z})
the mean square displacement
of FLRW can be shown as
\begin{equation}
\sigma(z)=g(z) e^{-\frac{2}{3+q}\frac{\gamma}{v}z},
\end{equation}
where $g(z)$ is the solution of Eq.
(\ref{Simplified first-order equation in main text}).
Since the mean square displacement $\sigma (z)$
monotonically increases with position $z$,
but $e^{-2 \gamma z /((3+q)v)}$
monotonically decreases with $z$,
so $g(z)$ should be the monotonically
increasing function of position
$z$. In what follows, we explore some special cases of Eq.
(\ref{Simplified first-order equation in main text}).

\subsection{The case $\boldsymbol{B=0}$}
For the case $B=0$ corresponding to magnetostatic
turbulence, Eq. (\ref{Simplified first-order
equation in main text}) can be simplified as
\begin{equation}
\frac{d^2g}{dz^2}=Ag^{-\frac{1+q}{2}}.
\label{eq:d2gdz2}
\end{equation}

By setting $g=\alpha |z|^\beta$ and using the
relationship $dg/dz=\alpha\beta|z|^{\beta-1}$ and
$ d^2g/dz^2=\alpha\beta(\beta-1)|z|^{\beta-2}$,
the Eq. (\ref{eq:d2gdz2}) can be rewritten as
\begin{equation}
\alpha\beta(\beta-1)|z|^{\beta-2}
=A\alpha^{-\frac{1+q}{2}}
|z|^{-\frac{1+q}{2}\beta}.
\label{Equation of alpha and beta}
\end{equation}

From Eq. (\ref{Equation of alpha and beta})
we can obtain the formulas of $\alpha$
and $\beta$ as follows
\begin{eqnarray}
\alpha &=& \left[\frac{(3+q)^2}{4(1-q)}\frac{D(s,q)}{2^{(1-q)/2}}
\frac{\delta B_{2D}^2}{B_0^2}\Gamma(\frac{q+1}{2})
l_{2D}^{1+q}\right]^{\frac{2}{3+q}},
\label{alpha}  \\
\beta &=& \frac{4}{3+q}.      \label{beta}
\end{eqnarray}

By combining Eqs. (\ref{alpha}) and (\ref{beta})
one can get the mean square displacement
of magnetostatic turbulence as
\begin{equation}
\sigma=\left[\frac{(3+q)^2}{(1-q)}\frac{D(s,q)}{2^{(5-q)/2}}
\frac{\delta B_{2D}^2}{B_0^2}\Gamma(\frac{q+1}{2})
l_{2D}^{1+q}\right]^{\frac{2}{3+q}}|z|^{\frac{4}{3+q}}.
\end{equation}

The latter formula is in agreement
with previous analytical result
derived in Shalchi (2011).

\subsection{The case $\boldsymbol{B=\varepsilon}$}
If $B$ is a very small quantity,
perturbation method (see, e.g., Paulsen 2014) can be used
to treat Eq.
(\ref{Simplified first-order equation in main text}).
Expanding $g(z)$ as a series of $\varepsilon$
\begin{equation}
g=g_0+\varepsilon g_1+\varepsilon^2 g_2
+\varepsilon ^3 g_3+ \cdots\cdots,
\label{Perturbation expansion}
\end{equation}
and inserting Eq. (\ref{Perturbation expansion})
into Eq.
(\ref{Simplified first-order equation in main text}),
one can achieve the following equations
\begin{eqnarray}
&&\varepsilon^0 : \frac{d^2 g_0}{dz^2}=Ag_0 ^{-\frac{1+q}{2}},
\label{Zero-order equation}  \\
&&\varepsilon^1 : \frac{d^2 g_1}{dz^2}-2\frac{d g_0}{dz}
=-\frac{1+q}{2}A\frac{g_1}{g_0^{(3+q)/2}},
\label{First-order equation}   \\
&&\varepsilon^2 : \frac{d^2 g_2}{dz^2}-2\frac{d g_1}{dz}
+g_0=Ag_0^{-(1+q)/2}\left[-\frac{1+q}{2}\frac{g_2}{g_0}
+\frac{(1+q)(3+q)}{8}\left(\frac{g_1}{g_0}\right)^2
\right]
\end{eqnarray}
and
\begin{equation}
\begin{aligned}
\varepsilon^3 : \frac{d^2 g_3}{dz^2}-2\frac{d g_2}{dz}+g_1
&=Ag_0^{-(1+q)/2}\Bigg[-\frac{1+q}{2}\frac{g_3}{g_0}
+\frac{(1+q)(3+q)}{4}\frac{g_1 g_2}{g_0}
-\frac{(1+q)(3+q)(5+q)}{48}\left(\frac{g_1}{g_0}\right)^3
\Bigg]
\end{aligned}
\end{equation}
and so on.

We can easily obtain the solution of Eq.
(\ref{Zero-order equation}) as
\begin{equation}
g_0 (z)=\left[\frac{(3+q)^2}{(1-q)}
\frac{D(s,q)}{2^{(5-q)/2}}\frac{\delta B_{2D}^2}
{B_0 ^2}\Gamma(\frac{q+1}{2})l_{2D}^{1+q}\right]^
{\frac{2}{3+q}}|z|^{\frac{4}{3+q}},
\label{Solution of zero-order equation}
\end{equation}
which corresponds to magnetostatic case.

The solution of Eq. (\ref{First-order equation})
can be found in Appendix,
here we directly show it as the following
\begin{eqnarray}
g_1&=&c_1 z^{2(1+q)/(3+q)}+c_2 z^{(1-q)/(3+q)}+b_0
z^{(7+q)/(3+q)}, \hspace{0.5cm}  \text{for}
\hspace{0.5cm} q\ne -1/3  \\
g_1&=&\left(c_1+c_2 lnz\right)\sqrt{z}+b_0 z^{(7+q)/(3+q)}.
\hspace{3.0cm}  \text{for} \hspace{0.5cm} q= -1/3
\end{eqnarray}
And using the same technique the
solutions of higher-order
equations can also be obtained successively.

By combining the solutions of all order
equations we can get the general solution
of Eq.
(\ref{Simplified first-order equation in main text})
as follow
\begin{equation}
\sigma=\left(g_0+\varepsilon g_1+\varepsilon ^2
g_2+\varepsilon ^3 g_3+\cdots\cdots\right)
e^{-\frac{2}{3+q}\frac{\gamma}{v}z}.
\label{Solution of Inhomogeneous equation for B=small value}
\end{equation}
For $\varepsilon =0$ we can get the solution
corresponding to magnetostatic turbulence.
And if $\varepsilon \ne 0$,
the terms $\varepsilon z^{(7+q)/(3+q)}, \cdots\cdots$
occur on the right-hand side of Eq.
(\ref{Solution of Inhomogeneous equation for B=small value}).
So MSD is a complicated function of $z$,
and only according to the formula
(\ref{Solution of Inhomogeneous equation for B=small value})
we cannot find the regimes of FLRW.
In what follows, the possible regimes
of FLRW and the influence of
temporal effect on the regimes will be explored.

For $\gamma=0$
corresponding to magnetostatic case, the power law exponent
$\beta$ is equal to $4/(3+q)$ which is the well known result
in diffusion theory of FLRW
(see Shalchi \& Kourakis, 2007a).
For the case that $\gamma$ tends to $\infty$,
it is more convenient to directly investigate
Eq. (\ref{Nonlinear Second order derivative equation of MSD}).
And if temporal factor $\gamma$ tends
to infinity
we can find that the right hand side of Eq.
(\ref{Nonlinear Second order derivative equation of MSD})
tends to zero. Therefore, for this case FLRW tends to diffusion, i.e.,
$\beta$ tends to 1.
In the following we explore the case of $\gamma \ne 0, \infty$.

Firstly, since the right-hand side
of Eq. (\ref{Nonlinear Second order derivative equation of MSD})
is greater than zero for $\gamma \ne 0, \infty$,
so the left-hand side is also greater than zero,
i.e., $d^2\sigma/dz ^2>0$.
So Eq. (\ref{Nonlinear Second order derivative equation of MSD})
only describe superdiffusive process.
Secondly, considering that $\sigma$ is the function
of temporal factor $\gamma$,
we explore the variation rule of $\sigma$ with $\gamma$.
After taking derivative of
Eq. (\ref{Nonlinear Second order derivative equation of MSD})
over $\gamma$ we can obtain
\begin{equation}
\frac{d^3\sigma}{dz^2d\gamma}=\frac{D(s,q)}{2}
\Gamma(\frac{q+1}{2})\frac{\delta B_{2D}^2}{B_0 ^2}
(2l_{2D}^2)^{(1+q)/2} e^{-\gamma z/v}\sigma^{-(3+q)/2}
\left(-\frac{z}{v}\sigma-
\frac{1+q}{2}\frac{d\sigma}{d\gamma}\right).
\end{equation}
Because $d^2\sigma/dz ^2>0$,
obviously, only $d\sigma/d\gamma<0$
satisfies the latter equation.
That is, temporal effect reduces MSD of FLRW.

To summarize the above discussion,
FLRW is superdiffusive for the case $\gamma<\infty$,
and it tends to diffusive
if $\gamma$ tends to infinity.
But subdiffusion does not occur.
So temporal effect could
change the regimes of FLRW
from superdiffusion into diffusion.
From Eq.
(\ref{Solution of Inhomogeneous equation for B=small
value}) we can find that MSD of FLRW
with dynamical effect is no longer the simple form
$\sigma=\alpha |z|^\beta$ or the logarithmic form
$\sigma\sim z lnz$ (see Kourakis et al. 2009).
Thus we can see that the temporal
effect not only changes the specific form
of MSD but also
affects the diffusion regimes of FLRW.
In fact, this is not a new result.
Some previous papers already have found
this effect (see Shalchi 2010a,
Guest \& Shalchi 2012).

\subsection{The condition of neglecting the term $\boldsymbol{Ag^{-(1+q)/2}}$
in equation $\boldsymbol{d^2g/dz^2-2Bdg/dz+B^2 g-Ag^{-(1+q)/2}=0}$}
Eq. (\ref{original Simplified first-order equation in main text})
can be nondimensionlized as the following
\begin{equation}
\frac{d^2g'}{dz'^2}-\frac{4}{3+q}\frac{dg'}{dz'}
+\left(\frac{2}{3+q}\right)^2 g'-\frac{1}{4}\frac{D(s,q)
\Gamma(\frac{q+1}{2})}{g'^{(1+q)/2}}R^2=0
\label{Inhomogeneous dimensionless second-order derivative equation}
\end{equation}
with $g'=g /(2l_{2D}^2)$, $z'=\gamma z/v$, and
$R=(\delta B_{2D} v)/(B_0 \gamma l_{2D})$.
Here we can see that the dimensionless quantity $R$
controls the physical process described by the
latter equation. If dimensionless quantity
$R$ is mall enough or $g'=g/(2l_{2D}^2)$
is large enough, the inhomogeneous equation
can be simplified down to homogeneous equation.
In what follows, we start from Eq.
(\ref{Simplified first-order equation in main text})
to explore the condition that
the term $Ag^{-(1+q)/2}$ can be ignored.

If the fourth term $Ag^{-(1+q)/2}$ on the left-hand
side of Eq.
(\ref{Simplified first-order equation in main text})
is much less than the other terms $d^2g/dz^2, 2Bdg/dz$
and $B^2 g$, the term $Ag^{-(1+q)/2}$ can be neglected.
Then we can get
\begin{equation}
\frac{d^2g}{dz^2}-2B\frac{dg}{dz}+B^2 g=0.
\label{Homogeneous Second-order equation with B}
\end{equation}
The general solution of the latter equation
can be obtained as
\begin{equation}
g=(c_1+c_2 z)e^{Bz}.
\label{Solution of Homogeneous Second-order equation with B}
\end{equation}
Then MSD can be written as
\begin{equation}
\sigma=ge^f=c_1+c_2 z.
\label{MSD for Homogeneous Second-order equation with B}
\end{equation}
If we set $\sigma(z=0)=0$, $c_1$ is equal to zero.
Since $\sigma(z)$ is positive, parameter $c_2$
also should be positive.
From the latter equation we can see that
Eq. (\ref{Homogeneous Second-order equation with B})
describes diffusion process.

By employing
Eq.
(\ref{Solution of Homogeneous Second-order equation with B})
the formulae $d^2g/dz^2, 2Bdg/dz$ and $B^2 g$ in
Eq. (\ref{Homogeneous Second-order equation with B})
can be obtained as follows
\begin{eqnarray}
\frac{d^2 g}{dz^2}&=&2c_2 B e^{Bz} +(c_1 +c_2 z)B^2 e^{Bz},
\label{first term }   \\
2B\frac{dg}{dz}&=&2c_2 Be^{Bz}+2(c_1+c_2 z)B^2 e^{Bz},
\label{Second term}  \\
B^2 g&=& (c_1 +c_2 z)B^2e^{Bz}.
\label{Third term}
\end{eqnarray}
Comparing Eqs. (\ref{first term })-(\ref{Third term}),
since $c_1 =0$ and $c_2>0$ and $B>0$ we can find that
$B^2 g$ is smaller than $d^2g/dz^2$, $2Bdg/dz$.
Therefore, if the
term $A g^{-(1+q)/2}$ in Eq.
(\ref{Simplified first-order equation in main text})
could be neglected,
the following condition should be satisfied
\begin{equation}
B^2 g\gg A g^{-(1+q)/2}.     \label{Raw inequality}
\end{equation}
By using formula $\sigma=g(z) e^{f(z)}$
we can rewrite Eq. (\ref{Raw inequality}) as
\begin{equation}
\sigma \gg \left(\frac{A}{B^2}\right)^{2/(3+q)}e^f,
\label{Raw inequality-2}
\end{equation}
which is the condition that the term ${Ag^{-(1+q)/2}}$
in equation ${d^2g/dz^2-2Bdg/dz+B^2 g-Ag^{-(1+q)/2}=0}$
can be neglected.

\subsection{Summary for $\boldsymbol{-1<q<1}$ in the range $\boldsymbol{2l_{2D}^2\ll \sigma\ll 2L_{2D}^2}$}
In this section we explore the properties of
FLRW for $-1<q<1$ in the range $2l_{2D}^2 \ll \sigma \ll 2L_{2D}^2$.
From Eq. (\ref{Raw inequality-2})
the following formula can be obtained
\begin{equation}
\left(\frac{A}{B^2}\right)^{2/(3+q)}e^f \ll \sigma \ll 2L_{2D}^2.
\end{equation}
By combining Eqs. (\ref{A}) and (\ref{B})
the latter formula can be written as
\begin{equation}
\left[\frac{D(s,q)}{2^{(1-q)/2}}\frac{\delta B_{2D}^2}
{B_0 ^2}\frac{v^2}{\gamma ^2}\Gamma(\frac{q+1}{2})
l_{2D}^{1+q}\right]^{\frac{2}{3+q}}\left(\frac{3+q}{2}\right)
^{\frac{4}{3+q}}\ll 2L_{2D}^2 e^{\frac{2}{3+q}\frac{\gamma}{v}z}.
\label{inequality}
\end{equation}
By using nondimensionlizing method
formula ({\ref{inequality}}) can be rewritten as,
\begin{equation}
\left(\frac{3+q}{4}\right)^{\frac{4}{3+q}}
\left[D(s,q)\Gamma(\frac{q+1}{2})\right]^{\frac{2}{3+q}}
R^{\frac{4}{3+q}}\xi^2\ll e^{\frac{2}{3+q}\chi},
\label{dimensionless inequality}
\end{equation}
where we use the following dimensionless quantities
\begin{equation}
\chi=\frac{\gamma}{v}L_{2D}, \hspace{0.5cm}
\xi=\frac{l_{2D}}{L_{2D}}, \hspace{0.5cm}
R=\frac{\delta B_{2D}}{B_0}\frac{v}{\gamma l_{2D}}.
\end{equation}

Obviously, from the latter inequality
we can see that the dimensionless quantities $R$,
$\xi$ and $\chi$ determine whether the term on
right-hand side of Eq.
(\ref{Simplified first-order equation in main text})
can be ignored or not.
When dimensionless quantities $\chi$
and $\xi$ are assumed constants and
$R$ is small enough, inequality
(\ref{dimensionless inequality})
could hold. Similarly, when $\chi$
and $R$ are fixed and
$\xi$ is small enough, or when $\xi$
and $R$ are fixed and $\chi$ is
large enough, inequality (\ref{dimensionless inequality})
also could hold.

In the previous paragraphs by using nondimensionlizing method
we explore the condition
that inequality (\ref{dimensionless inequality}) holds.
In what follows, we investigate the same issue
by directly using the pertinent physical quantities in real space.
Let us suppose that the inertial range index $s$,
the energy range index $q$, bend-over scale $l_{2D}$,
box scale $L_{2D}$， and turbulence level $\delta B_{2D}/B_0$
are all constants, and if temporal factor $\gamma$
is large enough,
inequality (\ref{dimensionless inequality}) could hold.
If we set the quantities
$s$, $q$, $l_{2D}$, $L_{2D}$, and  $\gamma$ as constants,
inequality (\ref{dimensionless inequality})
is established as long as $\delta B_{2D}/B_0$ is
sufficiently small. On the other hand, if the quantities
$s$, $q$, $L_{2D}$, $\delta B_{2D}/B_0$, and $\gamma$ are
constants, but the bend-over scale $l_{2D}$ is small enough,
or the quantities $s$, $q$, $l_{2D}$, $\delta B_{2D}/B_0$,
and $\gamma$ are constants, but the box scale is large enough,
inequality (\ref{dimensionless inequality}) could
also be established.
Therefore, not only the energy index $q$ and temporal
effect could change the regimes of FLRW,
but also the bendover scale $l_{2D}$, box size $L_{2D}$,
turbulence
level $\delta B_{2D}/B_0$
also could affect the regimes of FLRW.
In fact, for a certain magnetic turbulence system
the physical quantities
$s$, $q$, $l_{2D}$, $L_{2D}$, $\delta B_{2D}/B_0$, and $\gamma$
are probably dependent of each other.
So the number of quantities that
could affect independently the regimes of FLRW
might be less than $6$.

\section{SUMMARY AND CONCLUSION}
In the present article, by using field line tracing method
we have investigated analytically the properties of FLRW
in all possible length scales.
For simplification only
the simple damping dynamical model $\Gamma(\vec{k},t)=e^{-\gamma t}$
with constant factor $\gamma$ is employed.
The model $\Gamma(\vec{k},t)=e^{-\gamma t}$
denotes that the temporal correlation function
decay exponentially.
Of course, as a matter of fact,
temporal factor $\gamma$ should be the function of wave number $k$
and Alfv\'en speed $v_A$ and so on
(see, e.g., Shalchi 2010a, Guest \& Shalchi 2012).
For the purpose of simplification in this article
we assume temporal factor $\gamma$ is a constant
and leave the variable factor $\gamma$ for the future task.
By the investigation and discussion in this article
we find that if temporal effect is strong enough %$\gamma$
it can affect the field
line wandering in the range $\sigma\ll 2l_{2D}^2\ll 2L_{2D}^2$,
and the energy range spectral index $q$ determines
the properties of field line wandering
in the range $2l_{2D}^2\ll 2L_{2D}^2\ll\sigma $ and for the case $q>1$
in the range $2l_{2D}^2\ll\sigma\ll 2L_{2D}^2$,
but for the case $-1<q<1$ in the range $2l_{2D}^2\ll\sigma\ll 2L_{2D}^2$
both $\gamma$ and $q$ influence
the field line wandering. In addition, we obtain the following results.

(1) In order to describe the properties of FLRW of pure 2D turbulence
with damping dynamical model a new dimensionless
parameter $ R=(\delta B_{2D} v)/(B_0 \gamma l_{2D})$ is needed to be
introduced. For pure 2D turbulence Kubo number does not exist,
but $R$ does.

(2) If the temporal effect is strong enough,
it could change the transport regimes from
ballistic process into diffusive one in the range
$\sigma\ll 2l_{2D}^2\ll 2L_{2D}^2$.
In the range $2l_{2D}^2\ll\sigma\ll 2L_{2D}^2$
the temporal effect could change the transport regimes
from superdiffusion into diffusion but without subdiffusion.
However, the temporal effect does not influence FLRW
in the range $2l_{2D}^2\ll 2L_{2D}^2\ll\sigma$
regardless of the strength of temporal effect.
That is, temporal effect has no any impact
on FLRW in the range outside box size.

(3) Dimensionless mean square displacement $\sigma'$
is in proportion to $R^2$ in the range
$\sigma\ll 2l_{2D}^2\ll 2L_{2D}^2$. And $\sigma'$ is
in linear relation with $R^{4/(3+q)}$
for the case $-1<q<1$ in the range
$2l_{2D}^2\ll\sigma\ll 2L_{2D}^2$. But $\sigma'$
and $R$ are in the direct ratio
for the case $q>1$ in the range
$2l_{2D}^2\ll\sigma\ll 2L_{2D}^2$ and
for any allowed value $q$ in the range
$2l_{2D}^2\ll 2L_{2D}^2\ll\sigma $. Of course,
there are the same relationship as listed
above between mean square
displacement $\sigma$ and turbulence level
$\delta B_{2D}/B_0$.
Although temporal effect can reduce
FLRW and even change the regimes of FLRW,
it does not affect the relationship
between dimensionless mean square displacement
$\sigma'$ and dimensionless quantity $R$
in all possible length scales.
That is, temporal effect does not
affect the relationship
between mean square displacement $\sigma$ and
turbulence level $\delta B_{2D}/B_0$.

The dimensionless quantity $R$ introduced in this article
is related to temporal effect of turbulence.
In the future we will use real space method
and the Unified
NonLinear Transport (UNLT) theory (see Shalchi 2010b)
and so on
to dig more deeply into the features of
the new dimensionless quantity $R$.

In this paper, we concentrate
on the effects of the temporal factor
and energy range index
on the regimes of field line wandering.
However, other physical
properties such as turbulence level, bend-over scale,
and box-scale might also have
their effects. In addition, we only
explore field line wandering for damping model with
constant temporal factor. But the temporal factor might
be the function of wave number and
Alfv\'en wave speed and so on
(see, e.g., Shalchi et al. 2007;
Shalchi 2010a; Guest \& Shalchi 2012).
Therefore, our results might be oversimplified.
Furthermore, the properties of the field
line wandering with other dynamical models,
e.g., plasma wave model,
sweeping damping model, etc,
are also important research topics.
Moreover, the influence of dynamical effect
on energetic charged particle's transport
is another key problem worth paying attention.
Finally, the dimensionless quantities
corresponding to the above physical problems
also need to be explored carefully.
We will explore the problems listed above
in the future work.

\acknowledgments

We are partly supported by
grants NNSFC 41125016,  NNSFC 41574172, and NNSFC 41374177.

\renewcommand{\theequation}{\Alph{section}-\arabic{equation}}
\setcounter{equation}{0}  % reset counter
\begin{appendices}
\section{Solving the first-order Eq. (\ref{First-order equation})}
Substituting Eq. (\ref{Solution of zero-order equation})
into Eq. (\ref{First-order equation}), we get
\begin{equation}
\frac{d^2g_1}{dz^2}+\frac{2(1-q^2)}{(3+q)^2}
g_1z^{-2}=\left[\frac{D(s,q)}{2}\frac{\delta B_{2D}^2}{B_0^2}
\Gamma(\frac{q+1}{2})\left(2l_{2D}^2\right)
^{(1+q)/2}\frac{(3+q)^2}{4(1-q)}\right]^{2/(3+q)}
\frac{8}{3+q}z^{(1-q)/(3+q)}.
\end{equation}

The latter equation can be rewritten as
\begin{equation}
\frac{d^2g_1}{dz^2}+Mz^{-2}g_1=N z^{(1-q)/(3+q)},
\label{Simplified first-order equation}
\end{equation}
where we introduce the parameters
\begin{equation}
M=\frac{2(1-q^2)}{(3+q)^2},
\end{equation}
\begin{equation}
N=\left[\frac{D(s,q)}{2}\frac{\delta B_{2D}^2}{B_0^2}
\Gamma(\frac{q+1}{2})\left(2l_{2D}^2\right)^{(1+q)/2}
\frac{(3+q)^2}{4(1-q)}\right]^{2/(3+q)}\frac{8}{3+q}.
\end{equation}

Eq. (\ref{Simplified first-order equation})
is a Euler equation and the solution for $q \neq -1/3$ is as
\begin{equation}
g_1=c_1 e^{r_1 t}+c_2 e^{r_2 t}+b_0 e^{t(7+q)/(3+q)}
=c_1 z^{r_1}+c_2 z^{r_2}+b_0 z^{(7+q)/(3+q)},
\end{equation}
here the following parameters are used
\begin{eqnarray}
b_0&=&\frac{N}{\left(\frac{7+q}{3+q}\right)^2
-\left(\frac{7+q}{3+q}\right)+M}, \\
\end{eqnarray}
and for $q>-1/3$
\begin{eqnarray}
r_1&=&\frac{2(1+q)}{3+q}, \\
r_2&=&\frac{1-q}{3+q}
\end{eqnarray}
for $-1<q<-1/3$
\begin{eqnarray}
r_1&=&\frac{1-q}{3+q}, \\
r_2&=&\frac{2(1+q)}{3+q}.
\end{eqnarray}

And for $q=-1/3$ we can get the general solution
\begin{equation}
g_1=(c_1+c_2 lnz)\sqrt{z}+b_0 z^{(7+q)/(3+q)}.
\end{equation}

The specific expressions of $c_1$, $c_2$ can
be determined by using the specific conditions,
e.g., initial conditions, etc.
For simplification we only consider the cases for the energy index
$-1<q<1$ in the above discussion
because for $q>1$ some complicated cases of multiple roots occur.
\end{appendices}

{}

\end{document}